# ARTICLE

# Accelerated automated screening of viscous graphene suspensions with various surfactants for optimal electrical conductivity

Daniil Bash[a,b], Frederick Hubert Chenardi[a], Zekun Ren[c], Jayce Cheng[b], Tonio Buonassisi[c,d], Ricardo Oliveira[e], Jatin Kumar[b], Kedar Hippalgaonkar[b,f,#]

Functional composite thin films have a wide variety of applications in flexible and/or electronic devices, telecommunications and multifunctional emerging coatings. Rapid screening of their properties is a challenging task, especially with multiple components defining the targeted properties. In this work we present a manifold for accelerated automated screening of viscous graphene suspensions for optimal electrical conductivity. Using Opentrons OT2 robotic auto-pipettor, we tested 3 most industrially significant surfactants - PVP, SDS and T80 - by fabricating 288 samples of graphene suspensions in aqueous hydroxypropylmethylcellulose. Enabled by our custom motorized 4-point probe measurement setup and computer vision algorithms, we then measured electrical conductivity of every sample using custom and identified that the highest performance is achieved for PVP-based samples, peaking at 10.4 mS/cm. The automation of the experimental procedure allowed us to perform majority of the experiments using robots, while involvement of human researcher was kept to minimum. Overall the experiment was completed in less than 18 hours, only 3 of which involved humans.

## Introduction

In recent years, printable flexible electronics have received increasing attention.[1–5] Specifically, conductive inks and paints attract scientists' attention, as these materials show a wide range of applications, for example, printed radio frequency identification devices (RFID)[6], surface heaters[7], wearable printable sensors[8,9], IoT devices[10], conductive fabrics[11,12], etc. The most widely used additive for manufacturing conductive inks and paints is graphene[10,12–18] due to its intrinsic high electrical and thermal conductivity[19–21], chemical and thermal stability[22–25], and a relatively low price. However, because it is so chemically inert, graphene faces challenges when being made into dispersions, as it has very low affinity to most solvents, and, therefore, tends to aggregate and precipitate from the solution.

A common way to overcome this challenge is the use of surfactants.[4,12] Surfactants help to separate graphene flakes from each other and prevent them from aggregating, while also significantly increasing their affinity to the solvent[26,27]. However, excessive quantities of surfactant drastically reduce electrical conductivity of the resulting device since individual graphene flakes are insulated by surfactant molecules and cannot form an efficient path for charge carriers to flow through[28]. Hence, finding an optimal type and concentration of the surfactant is a key challenge for the industry as these properties have critical impact on the performance of the final device, bill of materials (BOM), and overall cost of manufacturing. Several studies have been carried out to identify the best surfactants for graphene dispersion in various solvents[26,28–31].

Surfactants like sodium dodecylsulphate (SDS), polyvinylpyrrolidone (PVP) and tween 80 (T80), among several others, have been shown to have the most optimal properties for graphene dispersion[32–35]. Industrially, the optimization of graphene-surfactant formulation is of major importance, since it directly relates to the performance of the inks and coatings, and to the final product cost. Generally speaking, one would expect that higher the surfactant proportion, the better the graphene dispersion, especially in greener solvents like water. However, excessive use of the surfactant penalizes the final ink/coating performance by, for example, reducing its electrical conductivity. To our knowledge, very few studies have explored the parameter space of the graphene-surfactant mixes in full or performed detailed characterization of the conductivity profiles of the samples. This can partially be explained by the fact that traditional manual methods of sample preparation are inadequate for covering a large parameter space. Therefore, we believe it is necessary to develop a robust high-throughput method for automation of liquid sample preparation, drop-casting and thin film characterization. The use of robotics and automation allows for highly reproducible, systematic

[a.] National University of Singapore, Singapore.
[b.] Institute of Materials Research and Engineering, Agency for Science, Technology and Research (A*STAR), Singapore.
[c.] Singapore-MIT Alliance for Research and Technology (SMART), Singapore.
[d.] Massachusetts Institute of Technology, Boston, MA, USA.
[e.] 2D Materials Pte Ltd (2DM), Singapore.
[f.] Nanyang Technological University, SIngapore
[#] Corresponding author.

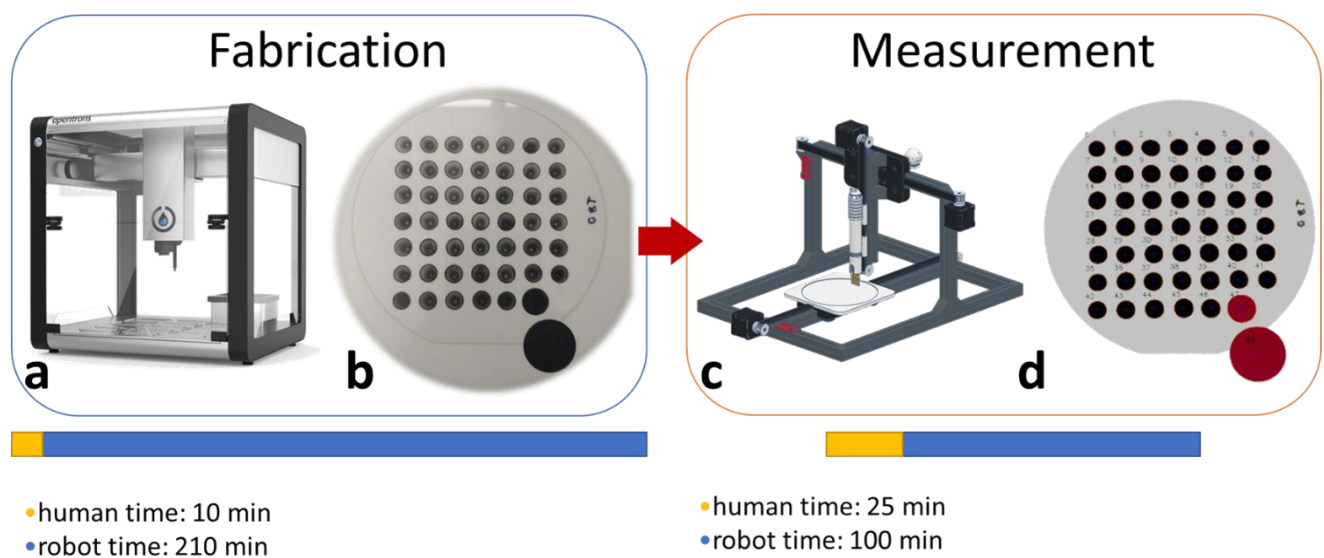

Figure 1. Schematic representation of the experimental workflow and distribution of time spent by humans and by robots. a) Fabrication of samples using Opentrons OT2 auto-pipettor (image credit: Opentrons); b) image of drop-casted samples; c) automated 4-point probe measurement setup; d) samples labelled by the computer-vision algorithm. Time estimations are provided per 96 samples, or one surfactant

fabrication of hundreds and thousands of samples with almost no human intervention.

In this work, we develop the methodology for fabrication of graphene-surfactant mixtures of various ratios in an automated fashion, thin film preparation and accelerated characterization. To achieve that, we used an Opentrons OT2 auto-pipettor (Figure 1a) to perform an exhaustive search of the full parameter space for ternary mixtures of graphene, hydroxypropylmethylcellulose (HPMC) and each of the 3 surfactants: PVP, SDS and T80. We used a custom python-based software to generate the design of experiment (DoE) in .csv format, which was used as instructions for the auto-pipettor for mixing and drop-casting.

An advantage of our approach is the robustness of the system for a varied range of viscous solutions. Specifically, we used HPMC, a common rheology modifier, to mimic the viscosity and the film-formation characteristic of an ink/paint[36]. Despite the added complexity of our control software, the auto-pipettor was able to handle viscous liquids without affecting the quality of the samples or the reproducibility of the experiment. Importantly, we significantly improved the efficiency and the throughput for fabrication of the samples, which was less than 2 minutes per sample, including mixing of the stock solution to obtain the desired graphene-surfactant ratio, and subsequent drop-casting, or approximately a 2-5-fold increase of throughput compared to manual process (Figure 1).

Rapid fabrication of test samples without a robust methodology for high-throughput characterization of these samples would have little benefit, especially for the use of supervised learning algorithms. Therefore, we have developed automated characterization techniques, which involve a 4-point probe for automated full IV measurement, as well as a computer vision-based algorithm for thickness approximation (Figure 1c and 1d respectively). The automation of the 4-point probe measurement has an added advantage of causing minimal damage to the samples and is highly reproducible due to the computer vision-based indexing of the samples' positions.

Traditionally, 4-point probe is pressed into the sample by hand and is held there by the researcher until the measurement is done, which takes a few minutes, in addition to positioning and setting up the instrument. However, it is almost impossible to hold the probe at exactly the same angle and at constant force during this procedure. Hence, the samples could be damaged by the probe slipping, and the results of measurement might be not consistent. In contrast, the automated manipulation of the probe allows to always apply uniform pressure and at exactly right angle, minimizing the damage to the sample and yielding highly reproducible results, while freeing up hands and mind of the researcher. The throughput was increased to approximately 1 minute per sample where a full wafer with 49 samples is automatically characterized in ~45 minutes including detection, alignment and measurement, which is 3-5 times faster than manual measurements (Figure 1c). The obtained IV curves were then converted into sheet resistance, which were used for further processing in this study.

The next automation step of our methodology is the computer vision-based thickness approximation algorithm. It was used for detecting the exact outline of the sample, calculated its area in pixels, which was then correlated to the true area of reference samples - 2 black circles with known area printed on A4 paper, placed underneath the wafer (Figure 1b and d, bottom right corner). Thereafter, we were able to calculate the thickness of the samples, based on known dispensed volume and concentration for every sample. This thickness data, combined with sheet resistance data, was used to calculate the property of interest – electrical conductivity. Overview of the experimental workflow s presented in a Figure 1 below. Overall, the full workflow, including fabrication and full characterization for all 288 samples, was completed in ~18 hours spread across 3 days, out of which ~15 hours were fully automated, therefore taking only 3 hours of focused human time. It is worth

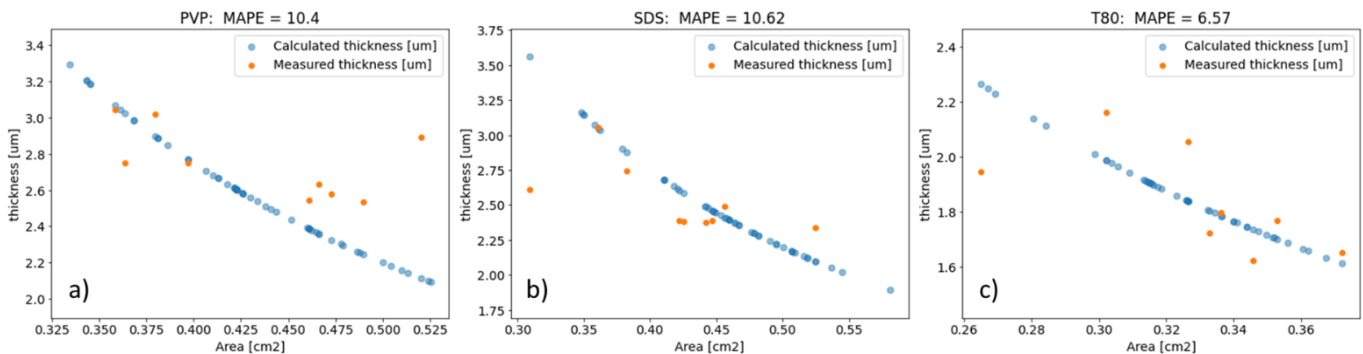
Figure 2. Comparison of calculated thickness and measured using profilometry. a) for PVP-based samples, b) for SDS-based samples, c) for T80-based samples. Each figure also shows value of the Mean Absolute Percentage Error (MAPE)

mentioning, that majority of the fabrication time was spent on programmed 3-second delays to let the viscous HPMC to flow in and out of pipette tip. When applied to non-viscous solutions, the same protocol is completed in under 10 minutes, compared to 3.5 hours for viscous ones. Further, because bulk of the experiment is done autonomously, the overall throughput of the experiment can be increased through parallelization of preparation of stock solutions, fabrication of the samples, and measurement of their properties.

This ability to perform experiments rapidly allows the use of dispersions that are stable over period of only few hours, which enables the researchers to broaden the parameter space and go beyond the compositions of infinitely stable dispersions. This opens a plethora of opportunities to explore large parameter spaces of increasing complexity in a high-throughput manner, enabled by a combination of automated mixing, drop-casting, and testing systems.

## Methods

The parameter space of graphene-to-surfactant ratios in HPMC was sampled uniformly with 2% increments with 8 to 18% surfactant for PVP and T80 samples, and 2 to 12% surfactant for SDS, where graphene was varied from 0 to 30% also with 2% increments, with solution of HPMC filling the remaining part of the composition. The higher initial ratio of surfactant for PVP and T80 was dictated by limitations of the stability of the stock solution of the graphene, while dispersions with SDS were comparatively stable at lower ratios of surfactant. The stock solutions of graphene were prepared by probe sonication for 3 hours of the mixture of graphene and surfactant at the highest graphene-to-surfactant ratio used for the given surfactant, so that the total concentration of solids was kept constant at 1 mg/ml. The stock solutions of HPMC and surfactants were prepared by simple mixing with deionized water at 1 mg/ml concentration and overnight stirring.

**Experimental setup**

The main tool used for fabrication of the samples in this work was the Opentrons OT2 auto-pipettor. Controlled by an onboard Raspberry Pi, it received instructions in the form of a .CSV file with volumes and positions for all operations. The robot then automatically distributed required amounts of ingredients into corresponding wells of a 96-well plate, using a new tip for each solution. After distribution, the robot was programmed to repeatedly aspirate and dispense the mixture in each well 10 times with varying heights to ensure proper mixing, and then drop-cast the resulting mixture onto a pre-treated fused silica wafer. The tip was washed by 5 repeated aspirations and dispensing of water each in 3 consecutive vials before moving to next sample.

While working with the HPMC solution, to account for its high viscosity, the robot was programmed to take a 3 second pause after every aspiration and dispensing operation, to allow the pressure in the pipette tip to stabilize. Another necessary added process step was the dipping of the pipette tip into the HPMC solution to a depth of <2mm from the meniscus, to minimize the HPMC's adhesion to the outer walls of the pipette and prevent interference with the determined dispense volume.

**Analytical setup**

After deposition and drying, the samples were tested for their electrical conductivity using custom automated 4-point probe measurement setup. The setup is equipped with a computer-controlled XY-stage, the probe itself, which is connected to a Keithley 2450, and an optical camera. Prior to measurements, the camera takes a picture of the wafer with samples, to run an image recognition algorithm. This algorithm detects the exact contour of every sample, calculating its area in pixels and coordinates of the center of mass, which is then used to index the positions for the electrical measurements, as seen in Figure 1d. For each sample we performed a current-voltage (IV) curve, slope of which was then used to calculate the sheet resistance. All measurements were done in 4-point probe configuration, to ensure maximum quality of the data and a linear slope. After the sheet resistance was recorded, we calculated the true area of each sample by relating the measured area in pixels to an area of 2 reference samples of known sizes (Equations 1 and 2). Here A is area of the sample, and k is a calibration coefficient

$$A_{true} = k A_{pixel} \quad (1)$$

$$k = \frac{1}{2}\left(\frac{A^1_{ref}}{A^1_{pixel}} + \frac{A^2_{ref}}{A^2_{pixel}}\right) \quad (2)$$

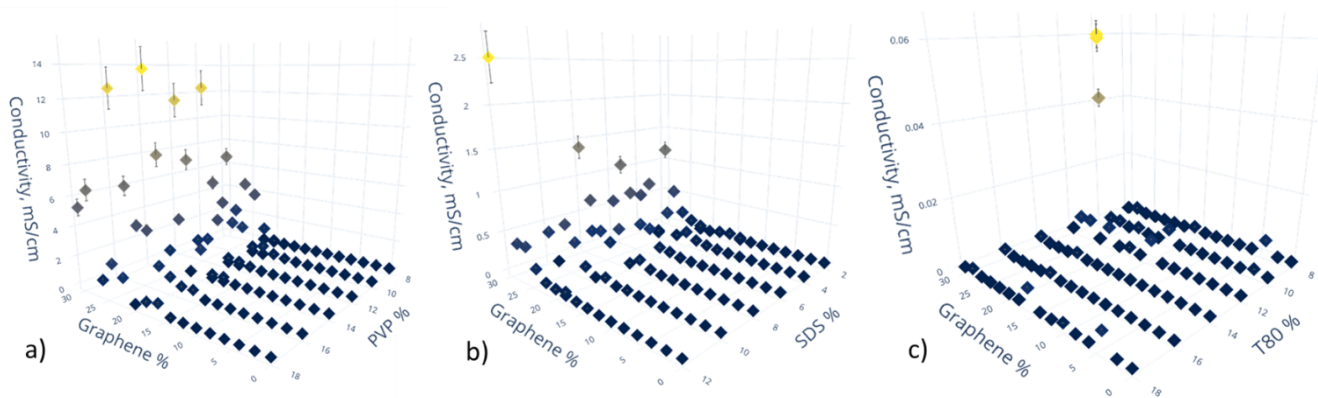

Figure 3. Conductivity in mS/cm vs graphene ratio in % vs surfactant ratio in %. Here, the percolation threshold is seen to be ~15% for PVP and ~20% for SDS, while it is not reached for T80. The conductivity is highest for the PVP, followed by SDS, while the Graphene-T80 film doesn't conduct.

The obtained coefficients were averaged and applied to the rest of the samples on the wafer to obtain the value of a true area. Then, combining it with known dispensed volume of each sample, we calculated the thickness of the samples. To make this link a few key assumptions are made: 1) the volume of the drop-casted samples is identical for all samples, down to technical limitations of Opentrons; 2) every sample has identical total concentration of solids, down to technical limitations of analytical balances and neglecting minute errors during transference of solutions; 3) the surfaces of the samples are flat and uniform, acknowledging that this assumption is the biggest source of error, but still accurate enough to be used for screening purposes. It is worth mentioning that calculated densities of all samples were found to be identical down to second significant decimal point. We then compared our calculated thickness data to those of randomly selected 7% of samples, measured on the surface profiler. Note that the surface profilometry has its intrinsic errors, since it requires a manual 'scratching' in the center of the sample and is by nature a single-line measurement, which is then extrapolated to the whole film. The mean absolute percentage error (MAPE) between the calculated thickness using the above assumption and measured thickness were found to be within 7-11% (Figure 3), hence supporting the feasibility of this method for high-throughput screening of thickness for large quantities of electronic composites.

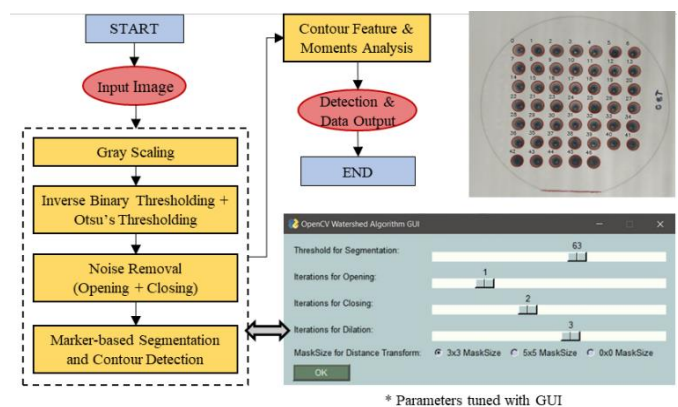

Figure 4. Schematic representation of the workflow of the computer vision algorithm (left), as well as the GUI for it (right)

An integral component of the optical thickness measurement technique is the implementation of computer vision-based (OpenCV) algorithms. The OpenCV pipeline consists of a chain of algorithms including segmentation and contour detection as well as data processing of the individual sample area, as seen in Figure 4. A graphical user interface (GUI) was built in Python using the PySimpleGUI library, to allow flexible, real-time tuning of the key parameters, such as segmentation threshold value, noise removal iterations, and distance transform mask size, to optimize the work of these algorithms.

Following segmentation and detection of the films, each sample is index-labelled with an integer value, which is sorted from top-left to bottom-right. The contour features and moments of each film are then analyzed to produce the area data in pixels.

Based on the area pixel values, the thickness of given film can be calculated by relating to the pixel values of 2 reference samples of known sizes (see Equations 1 and 2). To quantify the error between the calculated thickness and thickness obtained from surface profiler, we fit the calculated thickness to the area data and compute the MAPE for thickness values between the fitted sample and profiometry measured samples with different surfactants (PVP-based, SDS-based , and T80-based ). As shown in Figure 3, for all 3 cases, the MAPE is below 11% which is acceptable for the high throughput proxy measurement. We use a Gaussian Process (GP) model to reproduce the equation that relates area to thickness values. The GP model is chosen as it can account for potential white noise errors that could arise from the computer vison algorithm and gives an uncertainty estimate. The GP kernel implemented here consists of a Radial Basis Function (RBF) kernel and a Whitenoise Kernel.

## Results and discussion

The mechanism of the conductivity in graphene-based composites is typically explained by percolation theory[37–40], according to which individual sheets of graphene form a connected network, allowing the flow of the charge carriers. However, when the concentration of graphene is low, below a percolation threshold, a fully connected network cannot be formed, and the charge carriers hop between islands of graphene clusters via quantum tunneling, which significantly decreases the conductivity of the composite.

The results of our study show that films made using PVP as surfactant are the most conductive, with highest value of conductivity of 10.8 mS/cm, as seen in Figure 3a. Interestingly, conductivity of PVP-based films does not show dependance on the concentration of PVP, unlike films made with other surfactants. SDS-based films show low conductivity values for most of the tested conditions, except for the samples with high concentration of graphene, as seen in Figure 3b. Concentration of SDS showed no significant influence on the device performance. The worst conductivity values were observed for films made using T80, where the highest observed conductivity was only 0.06 mS/cm, as shown in Figure 3c, which is about 3 orders of magnitude lower than for PVP-based samples. Majority of the measured T80-based samples show conductivity values too low to be distinguished from instrument noise.

We speculate that PVP shows the best results due to higher affinity of its aromatic benzene rings to aromatic rings of graphene, as compared to other surfactants used[41]. Also, due to its polymer nature as compared to SDS, PVP could promote better alignment of graphene sheets and, therefore, increase the overall conductivity of the sample. SDS, on the other hand, despite showing mediocre conductivity results, could stabilize the dispersion at 4 times lower concentration compared to other tested surfactants. This feature could be significant for specific industrial applications, e.g. high conductivity ($10^{-1}$-$10^{2}$ S/cm) for paint-on sensors or low conductivity ($10^{-8}$-$10^{-4}$ S/cm) for anti-static coatings, as it could provide significant cost savings.

We explain the poor performance of the T80-based samples not by inherit incompatibility of this surfactant as stabilizing agent for graphene dispersions, but by excessively aggressive sonication parameters. The conditions used to sonicate the dispersions could have damaged the polymer chains to the point that the T80 lost most of its properties as a surfactant. The backbone of T80 is comprised of ester C-O bonds, which are weaker in nature, as contrasted to C-C bonds in PVP. Future experiments will aim to investigate this assumption, by using a series of more mild sonicating conditions with T80-based composites.

Further analysis of the surface of the composites using Scanning Electron Micrographs (SEM) revealed that the distribution of graphene inside the HPMC matrix is relatively uniform at lowest surfactant load and highest and medium graphene loads (Figure 5a-f and Figure 5g-l respectively), suggesting the effectiveness of our dispersion and fabrication procedures. The scanning electron microscopy (SEM) images reveal that the graphene islands inside the composites are relatively dense, however, to test how much impact the separation of these islands by HPMC inside the composite has on the performance of the composites, we decided to anneal the samples beyond the decomposition temperature of the HPMC and repeat the electrical measurements and SEM characterization.

Same samples used for SEM imaging of samples without any post processing were used to test the effect of annealing. The annealing was done in the vacuum furnace. First, the samples were brought from room temperature to 250 ºC at 10 ºC/min, and then from 250 ºC to 500 ºC at 1 ºC/min and held at 500 ºC for 120 min to completely burn off the HPMC binder.

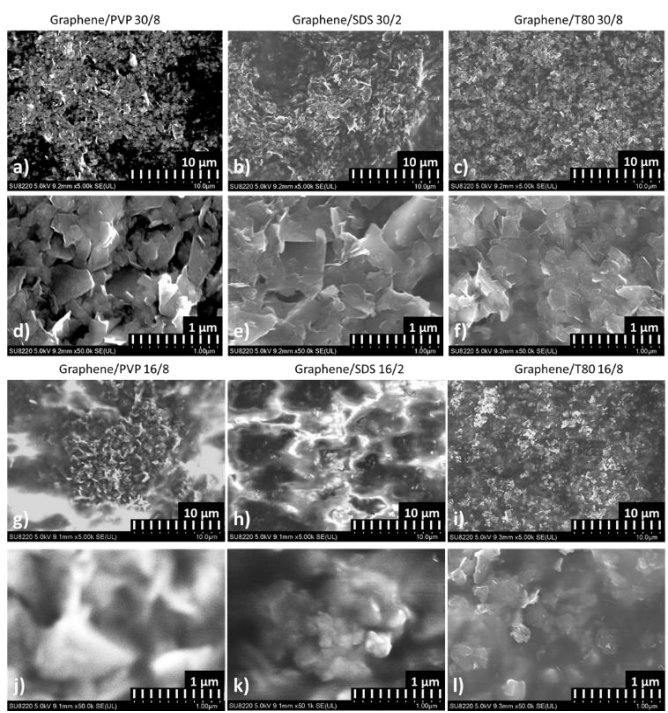

Figure 5. SEM images of untreated samples. a-c) 5k magnification of high graphene load samples; d-f) 50k magnification of high graphene load samples; g-i) 5k magnification of medium graphene load samples; j-l) 50k magnification of medium graphene load samples

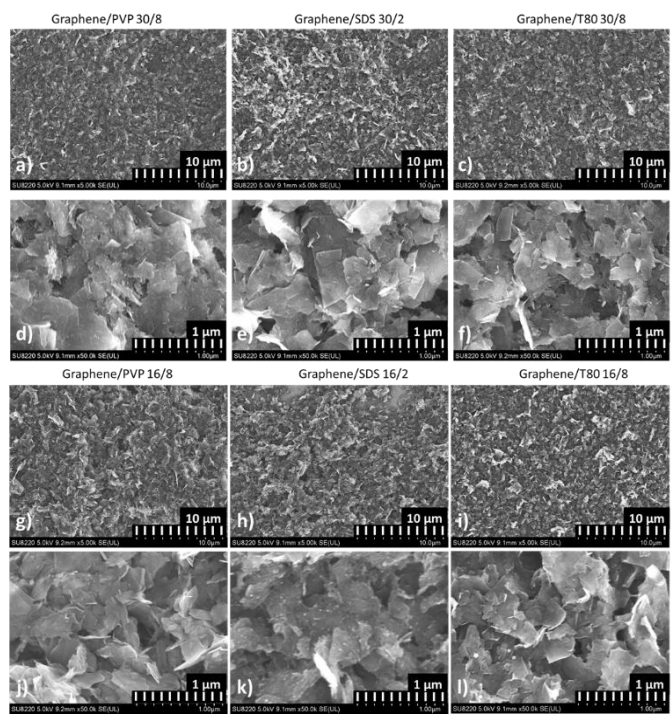

Figure 6. SEM images of untreated samples. a-c) 5k magnification of high graphene load samples; d-f) 50k magnification of high graphene load samples; g-i) 5k magnification of medium graphene load samples; j-l) 50k magnification of medium graphene load samples

After annealing, thickness and sheet resistance of the samples was measured again. We observed the conductivity increase up to 7 orders of magnitude, suggesting that with removed polymer matrix, graphene flakes got into close contact with each other to form a dense percolation network. The highest increase in conductivity is observed for, expectedly, T80-based samples, because they were barely conductive prior to annealing, while for SDS-based samples increase in conductivity is the lowest, most likely due to lower influence of annealing on stability of the surfactant molecule. As seen from the images in Figure 6, the morphology of the annealed samples does not seem to depend on the surfactant, nor on initial concentration of graphene inside the composite, as compared to untreated samples.

## Conclusions

In this work, we demonstrated a high-throughput methodology for automated mixing and drop-casting of graphene-based composites followed by their electrical conductivity. We used a robotic auto-pipettor to explore more combinations of parameters (288 unique conditions tested) in 10x less time, as compared to traditional procedures, even using viscous liquids. Distribution of all ingredients, mixing and subsequent drop-casting of all samples was done in under 12 hours, almost fully autonomously. In addition, when applied to electrical measurements, automation not only improves the throughput of the experiments, but also increases the reliability and reproducibility of the measurements, as every sample is precisely indexed and is measured with minimal damage to the film. We also show the utility of our method for rapid screening of thickness for thin-film samples, relying only on computer vision algorithms and precision of auto-pipettor to drop-cast exact volumes. Amongst all films prepared, the graphene dispersions made with PVP as surfactant have the highest conductivity values as compared to samples made with SDS and T80, with best result of 10.8 mS/cm. We also identified that the relative concentration of the surfactant plays minimal role in the overall performance of the composite, suggesting that the load of the graphene plays the most significant role.

In conclusion, using this approach we fabricated 288 samples in this study, with a goal to identify the best surfactant to be used for graphene dispersions, in less than a week of experimental work, most of which was done by robots. Hence, we demonstrate the viability and applicability of automation tools to scientific experiments, especially the ones which rely on many repetitive operations and exploration of vast parameter spaces. These techniques can free up most of human workhours from the experiments, and delegate the tedious work to robots. Our work demonstrates the push towards automation in science laboratories, where human researchers are engaged in creative scientific work and planning of experiments, while execution is delegated to robots, machine learning algorithms and efficient high-throughput experimentation and analytics tools.

## Author Contributions

Author contributions are provided in CRediT format.
D. Bash – conceptualization, investigation, methodology, formal analysis, software, validation, writing original draft, visualization;
F. H. Chenardi – software, writing original draft, visualization;
Z. Ren – formal analysis, visualization;
J. Cheng – software, methodology;
T. Buonassisi – conceptualization, project administration;
R. Oliveira – conceptualization, resources, writing review and editing;
J. Kumar – conceptualization, project administration, supervision;
K. Hippalgaonkar – conceptualization, project administration, supervision, funding acquisition, resources, writing review and editing.

## Conflicts of interest

The graphene sample for this study was provided by company 2DM free of charge.  Some of the authors own equity in companies applying machine learning to materials development.

## Acknowledgements

We'd like to thank Patrick Teyssoneyre for his effervescence and the members of the AMDM and Xinterra teams for discussions. We acknowledge funding from the Accelerated Materials Development for Manufacturing Program at A*STAR via the AME Programmatic Fund by the Agency for Science, Technology and Research under Grant No. A1898b0043.

## Notes and references